# Differences in Neurovascular Coupling in Patients with Major Depressive Disorder: Evidence from Simultaneous Resting-State EEG-fNIRS

Feng Yan#[1], Xiaobin Wang#[2,3], Yao Zhao[5], Shuyi Yang[4,6], Zhiren Wang[1]

1. Beijing Huilongguan Hospital, Peking University Huilongguan Clinical Medical School, Beijing, China
2. Tsinghua Laboratory of Brain and Intelligence
3. AI Innovation Center, Tsinghua Pearl River Delta Research Institute
4. School of Biomedical Engineering, Tsinghua university
5. Department of Psychology, School of Humanities and Social Sciences, Beijing Forestry University
6. Xinya College, Tsinghua University

# These authors contributed equally

**Abstract:**

Neurovascular coupling (NVC), the relationship between neural activity and cerebral hemodynamic responses, remains poorly understood in major depressive disorder (MDD). To investigate alterations in NVC associated with depressive symptom severity, we simultaneously recorded resting-state electroencephalography (rsEEG) and functional near-infrared spectroscopy (fNIRS) in 206 participants, including 134 patients with MDD and 72 healthy controls stratified by age to disentangle disease-related alterations from age-associated effects. NVC in the prefrontal cortex (PFC) was characterized by the consistency and temporal lag between spontaneous electrophysiological peaks and corresponding hemodynamic responses.

We found that age significantly enhances the NVC consistency ($p < 0.05$), whereas this relationship was altered in patients with MDD during both the oxygen-consumption phase and the subsequent hemodynamic compensation phase. Furthermore, among patients with MDD, the strength of this consistency decreased more significantly with greater illness severity (partial $r = -0.336$, $p = 0.0601$). These findings suggest that changes in neurovascular coupling effects are strongly associated with the severity of depression.

By leveraging wearable neuroimaging techniques, this study provides multimodal evidence for altered neurovascular coupling in depression and highlights its potential as a biomarker for disease monitoring and recovery trajectories.

**Keywords:**

Neurovascular coupling, Major Depressive Disorder, Disease severity, Resting-state, EEG-fNIRS

# 1. Introduction

Major depressive disorder (MDD) as a representative affective disorder currently affects over 280 million people worldwide, posing a significant public health challenge that demands urgent attention (Evans-Lacko et al., 2017). Over the past few decades, the diagnosis of MDD has progressed from relying solely on behavioral assessments and self-report questionnaires to incorporating objective neurophysiological and biological measurements. These include electroencephalography (EEG), which captures neural electrical activity; functional magnetic resonance imaging (fMRI), which reflects blood oxygen level-dependent (BOLD) signals; functional near-infrared spectroscopy (fNIRS), which similarly captures cerebral hemodynamic changes; wearable devices that monitor peripheral nervous system activity; and genetic sequencing.

For instance, EEG-based studies have revealed hemispheric asymmetry in alpha oscillations among individuals with depression, reporting reduced alpha power in the left frontal cortex relative to the right (Henriques & Davidson, 1991). Furthermore, increased delta (0.5–4 Hz) and theta (4–7 Hz) activity has been observed in the prefrontal regions of depressed patients (Jaworska et al., 2012). In parallel, neuroimaging techniques that measure cerebral hemodynamics—particularly fMRI—have provided valuable insights due to their high spatial resolution. A series of studies have demonstrated hyperactivity within the default mode network (DMN) in MDD, especially aberrant connectivity between the medial prefrontal cortex (mPFC) and the posterior cingulate cortex (PCC), which has been associated with ruminative thinking (Greicius et al., 2007; Pizzagalli & Roberts, 2022). Similar observations have been reported in fNIRS studies (Chao et al., 2021), collectively indicating that both neurophysiological and hemodynamic signals can effectively differentiate individuals with MDD from healthy controls.

To provide a more comprehensive description of the changes in the MDD, several studies have explored multimodal monitoring strategies, such as temporally separated EEG and fMRI acquisitions (Williams et al., 2011), or simultaneous acquisition of different modalities (e.g., EEG and ECG, or EEG and eye-tracking) (Williams et al., 2011; Zheng et al., 2018). These efforts aim to uncover inter-modal relationships to improve diagnostic precision. However, multimodal central nervous system data collection is difficult, largely due to the technical challenges of denoising. For instance, EEG-fMRI systems require MR-compatible equipment and face hurdles related to artifact correction and signal alignment, limiting their clinical scalability (Warbrick, 2022).

However, some neural mechanisms can only be elucidated through multimodal measurements. One such phenomenon is neurovascular coupling (NVC)—the process by which local neuronal activation leads to energy consumption, increased oxygen demand, vasodilation, cerebral blood flow (CBF) increase, and elevated oxygenated hemoglobin concentration (Phillips et al., 2016). Previous studies have shown that during oddball tasks, the hemodynamic response typically lags behind the neural response (e.g., P300) by approximately 1–2 seconds, peaking around 4–6 seconds post-stimulus (Kennan et al., 2002). The latency and morphology of the NVC response reflect vascular elasticity and can be influenced by aging, neurodegenerative disorders, and vasodilatory medications that act on smooth muscle cells (Phillips et al., 2016; Tarantini et al., 2017). While NVC has been extensively studied in the context of neurodegeneration, its relevance to psychiatric disorders such as MDD remains largely underexplored. A study utilizing EEG-fMRI investigated differences in NVC related to inhibitory function in late-onset MDD (Bondi et al., 2025). Beyond this, relevant research remains relatively limited overall. However, functional and structural brain changes have been observed in individuals with depression (Banasr et al., 2021; Zhang et al., 2018), suggesting that NVC alterations may be observed as potential biomarkers for disease progression from functional alterations to structural. Drawing upon insights from neurodegenerative research, assessing NVC in MDD may offer valuable information about how neural activity drives vascular adaptation. Moreover, it raises the question of whether vascular elasticity should be considered in the prognosis and recovery planning for severe mental illness.

In this study, we developed a plug-and-play EEG-fNIRS system capable of synchronously recording resting-state brain activity in the prefrontal cortex (PFC) to evaluate neurovascular coupling (NVC) in patients with major depressive disorder (MDD) across different stages of illness. To focus specifically on the MDD-related NVC phenomena, which target to the prefrontal cortex, all available fNIRS sources and detectors were systematically concentrated in this region, given the finite number of available optodes. To characterize the influence of aging on NVC and disentangle age-related effects from disease-related alterations, the healthy control groups were divided, and additional recruitment was conducted based on the age distribution of MDD patients for further analysis.

EEG was used to capture resting electrophysiological activity associated with neural activation, while fNIRS measured the corresponding hemodynamic responses. This approach reduces reliance on conventional task-based paradigms (e.g., oddball tasks), which assume relatively uniform evoked responses across individuals (Bondi et al., 2025; Kennan et al., 2002). Instead, it enables real-time monitoring of each participant's intrinsic neural activity and its associated vascular responses under resting-state conditions.

For the first time, this study investigates the neurovascular bases of resting-state in patient with MDD with different severity, compared to healthy controls, through wearable device. We aim to investigate the dynamic change of time-window characteristics in neurovascular coupling. We hypothesize that MDD patients will exhibit abnormal hemodynamic response patterns compared to healthy controls in brain regions involved in the modulation of instantaneous neural activity during resting state. These changes are expected to highlight both commonalities and distinctions in the neural mechanisms underlying personal state between MDD and HC, thereby advancing our understanding of depression-related dysfunctions.

**2.2 Participants**

A total of 206 participants were recruited for this study, including 134 patients with MDD, and 72 healthy controls.

All patients met the Diagnostic and Statistical Manual of Mental Disorders, Fifth Edition (DSM-5) criteria for major depressive disorder (MDD) and were at least in the acute, maintenance, or stable phase (based on ICD-10). This means that the study excluded participants who had a past history of depression but had already achieved full clinical remission at the time of enrollment. Healthy controls were screened to exclude any history of neurological or psychiatric disorders.

To better differentiate the severity levels of patients, we divided the patients into two groups based on whether their HAMD depression scale scores were greater than 16. *Table 1* shows that there were differences in mean age between the patient groups after stratification, which guided us to recruit additional healthy controls in a more targeted way.

**Table 1.**
Clinical details for participants.

| Groups | Gender | Age (mean±SD) | HAMD (mean±SD) | HAMA (mean±SD) | BDI (mean±SD) | SAS (mean±SD) | Medication |
|---|---|---|---|---|---|---|---|
| MDD (HAMD>16) | 16M, 42F | 23.34±9.59 | 23.84±5.54 | 25.78±7.27 | 24.00±10.28 | 56.18±10.46 | SSRI:32; SNRI:15 Others:11 (None/CBT:2) |
| MDD(HAMD ≤16) | 21M,55F | 32.14±10.28 | 8.54±4.19 | 10.49±6.98 | 8.09±6.61 | 41.86±8.51 | SSRI:35 SNRI:20 Others:21 (None/CBT: 6) |
| Healthy Controls | 22M,50F | 26.03 ± 6.81 | / | / | / | / | / |

AG: Acute phase group; MG: Maintenance phase group; HC: Healthy Controls group
BDI: the Beck Depression Inventory (BDI)
SAS: the Self-Rating Anxiety Scale (SAS)
HAMD: the Hamilton Depression Rating Scale (HAMD-17)
HAMA: the Hamilton Anxiety Rating Scale (HAMA)
SSRI: Selective Serotonin Reuptake Inhibitor
SNRI: Serotonin-Norepinephrine Reuptake Inhibitor

Each participant underwent a sequence of resting-state recordings: 2 minutes with eyes open, followed by 2 minutes with eyes closed. This sequence was then repeated to obtain two full cycles (open–closed–open–closed), ensuring sufficient relaxation and physiological stabilization. Participants were instructed to press a button to advance to each subsequent stage, and an auditory cue was presented to signal the end of each eyes-closed interval. It should be noted that some participants performed additional task paradigms during the rest intervals for other experiments(Yan et al., 2026).

The entire experiment took approximately 25 - 50 minutes. The study was approved by the Institutional Medicine Review Board of Tsinghua University (Approval number: *THU01-20240104*) and the Institutional Medicine Review Board of Beijing Huilongguan Hospital (approval number: 2024-106-research, 2026-10-research). All participants signed informed consent forms before the commencement of the experiment.

**2.3 EEG & NIRS Measurements：**

EEG signals were recorded using a wireless EEG acquisition system (NeuSen.W32, Neuracle, China) at a sampling rate of 1000 Hz. Thirty-two saline-based sponge electrodes were positioned according to the international 10–20 system. Electrode impedance was maintained below 30 kΩ throughout the experiment to ensure acceptable quality signal acquisition.

fNIRS signals were simultaneously recorded using a wireless near-infrared spectroscopy system (NirSmart, Danyang Huichuang Medical Equipment Co., Ltd., China) with a sampling rate of 11 Hz. A total of 51 channels, consisting of 16 source and 20 detectors, were arranged over the prefrontal and temporal cortices. The regions were specifically targeted due to their well-documented associations with abnormal cognitive processing in individuals with MDD (Zhu et al., 2017).

To ensure reliable contact between the EEG sponge electrodes and the fNIRS optodes with the scalp, we redesigned the attachment mechanism for the fNIRS optodes. Custom fixtures (white clips in the figure) were fabricated using 3D printing to enhance device stability and signal quality during recording. This modular and portable design not only reduced the setup time and improved user comfort but also provided a practical solution for scaling up future multimodal neuroimaging studies involving simultaneous EEG and fNIRS measurements.

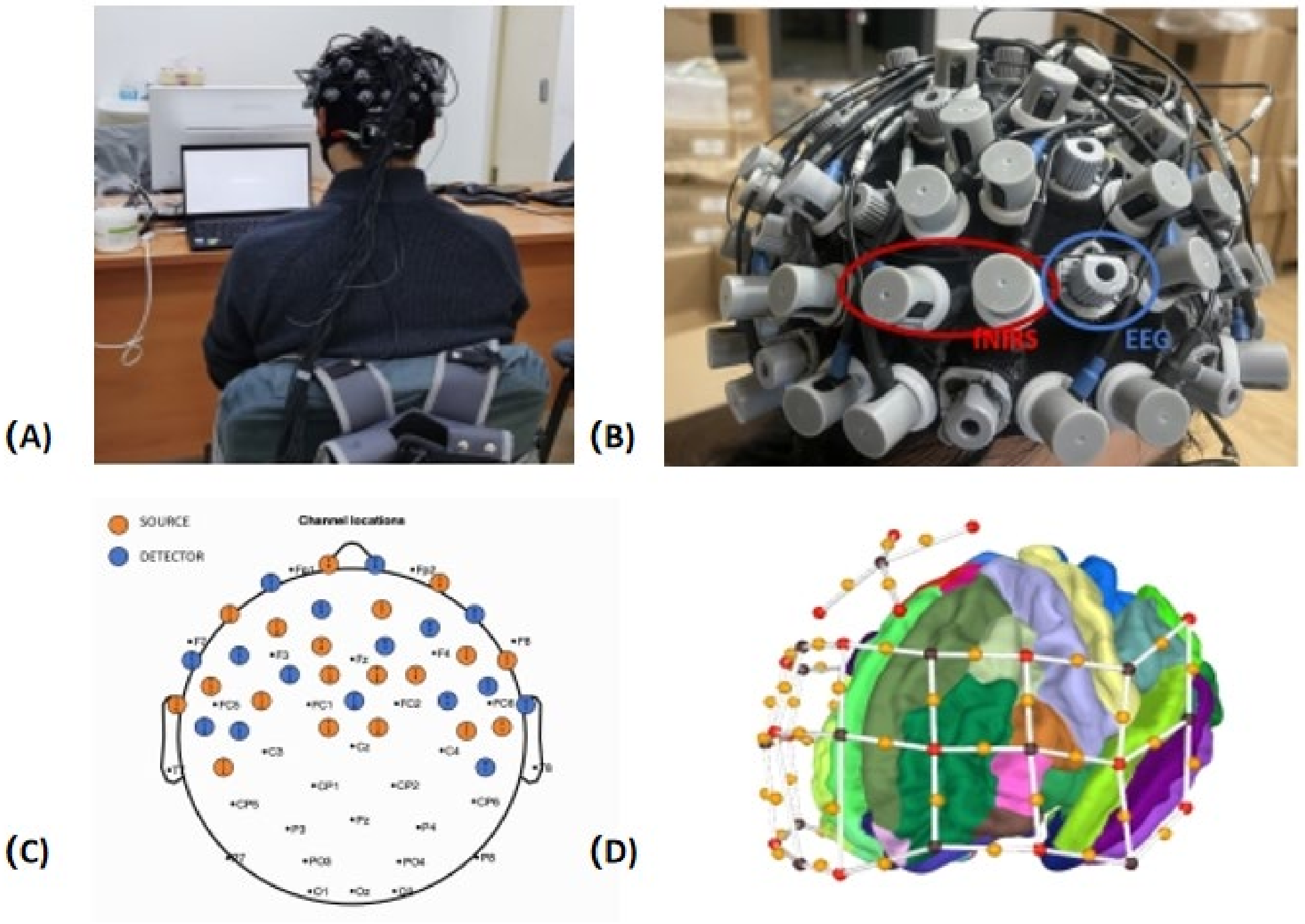

Fig.1. (A) and (B) depict the customized cap design. In (B), red circles indicate the positions of fNIRS optodes, while blue circles mark the saline-based sponge EEG electrodes. (C) shows the spatial distribution of fNIRS sources and detectors based on the standard 10-20 EEG system, with orange dots representing sources and blue dots representing detectors. (D) presents the 3D layout of fNIRS measurement channels, where red dots indicate sources, black dots indicate detectors, and yellow dots represent the theoretical measurement locations at the midpoint between each source-detector pair. Some source-detector distances are slightly shorter or longer than the standard 3 cm due to spatial constraints imposed by the EEG electrode positions. These variations were accounted for during fNIRS signal processing.

**2.4 EEG Preprocessing**

The EEG preprocessing pipeline was implemented using the EEGLAB toolbox (Version 2024.2.1). First, spectral analysis (0-60Hz range) was performed to visually inspect raw signals, with particular attention to identifying 50Hz line noise in T7/T8 channels resembling the artifacts previously observed in A1/A2 channels.

Subsequently, A1/A2 electrodes were excluded prior to executing standard preprocessing steps including filtering, bad channels interpolation, downsampling, epoch segmentation, artifact rejection through independent component analysis (ICA) for removing motion-related and ocular artifacts. Specifically, a band-pass filter (0.5-40 Hz) was applied using EEGLAB's *pop_eegfiltnew* function. Automated bad channel detection was conducted through *kurtosis_thresholding* (threshold = 5), with results cross-validated against manual spectral inspection. Identified problematic channels were reconstructed via spherical interpolation.

Following this, data were resampled to 250 Hz using the *pop_resample* function. Continuous data were then segmented into extended epochs to eye-open/eye-close event markers. Independent component analysis decomposed the data into 30 principal components. Component classification was performed using the ICLabel classifier: neural-origin components (labeled "Brain") were retained, while components classified as muscular ("Muscle"), cardiac ("Other"), ocular ("Eye"), channel noise("Other"), or line noise ("Other") were systematically rejected.

**2.5 fNIRS Preprocessing**

fNIRS data preprocessing was performed using the MNE-NIRS toolbox (v0.9). First, event-based segmentation was applied to extract 120-second epochs following each eye-open or eye-closed condition onset. Bad channels were identified based on the raw intensity signals: a channel was marked as bad if its mean light intensity was below 0.4. Unlike EEG preprocessing, these bad channels were discarded rather than interpolated, as low-quality signals in fNIRS are typically caused by excessive hair coverage, and interpolating between bad channels may introduce misleading artifacts. If more than one half the channels were classified as bad, the subject was excluded from analysis. In practice, a substantial number of participants were removed at this stage, likely due to the presence of dense, black long hair and worse levels of sanitation among the sample population (Holmes et al., 2024).Employing an EEG cap electrode configuration rather than a headband-style setup may similarly lead to a decrease in data quality.

Next, the raw light intensity data were converted into optical density using the *mne.optical_density* function. A band-pass filter (0.05 – 0.7 Hz) was applied to remove physiological noise such as heartbeats and respiration. Temporal baseline shifts and spike artifacts were then corrected using the Temporal Derivative Distribution Repair (TDDR) method via the *temporal_derivative_distribution_repair* function (Fishburn et al., 2019). Finally, the *beer_lambert_law* was applied with a partial pathlength factor (PPF) of 6 to transform the optical density signals into hemoglobin concentration changes (HbO(Oxyhemoglobin) and HbR (Deoxyhemoglobin)). The total hemoglobin concentration (HbT, normally HbO + HbR) was used for further analyses.

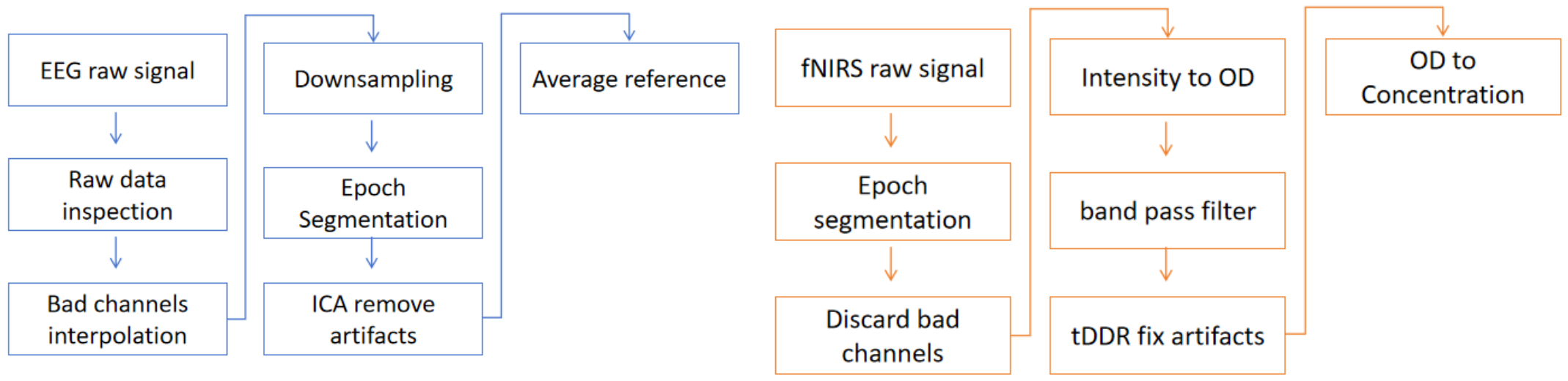

Fig.2. The preprocessing pipelines of EEG and fNIRS

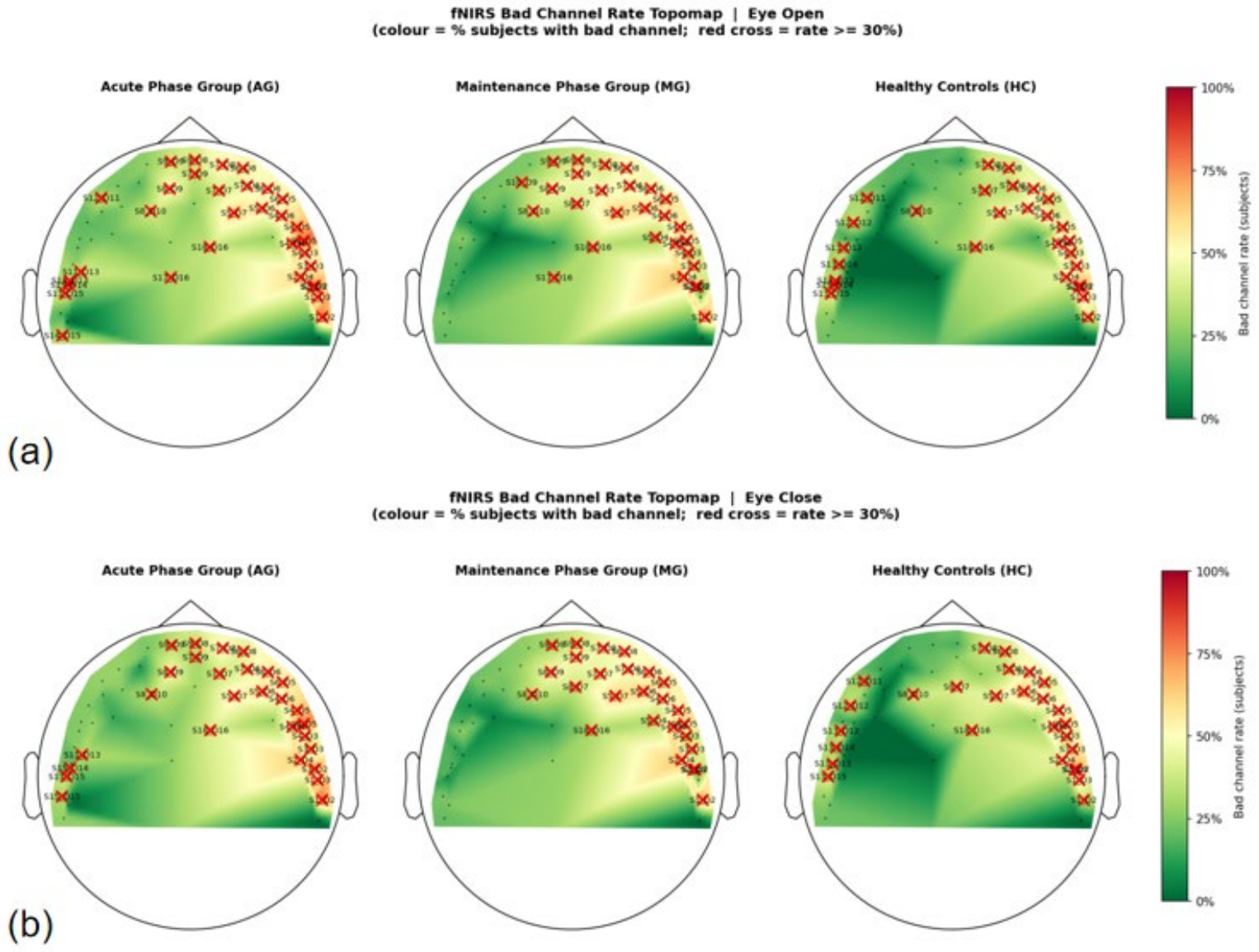

Fig.3. fNIRS channel rejection rates: The color scale indicates the proportion of rejected epochs at each location across subjects, with warmer colors representing higher rejection rates. Red 'X' markers denote channels where the rejection rate exceeded 30% of the total epochs. Panel (a) shows the Eyes-Open condition, and panel (b) shows the Eyes-Closed condition

Following rigorous data preprocessing and quality control, 74 participants were retained for NVC extraction and analysis. This high attrition rate (approximately 65%) was primarily attributed to the technical challenges of simultaneous EEG-fNIRS recording. Specifically, the presence of the EEG cap hindered optimal scalp-optode coupling; the subject's hair often obstructed near-infrared light paths, and the configuration made it difficult to adequately part the hair for stable contact, leading to significant signal loss and subsequent data exclusion. This situation was more pronounced in female participants with long or permed hair. Information regarding data quality across different fNIRS channels is shown in the *Fig.3*.

The demographic details of the participants included in the final NVC calculations are presented in *Table 2*. To better control for the effects of age in the healthy control group, we divided them into two subgroups using an age cutoff of 28. This resulted in 17 participants in 'Healthy Group 2'—a size comparable to the retained MDD patient in the HAMD<= 16 group, with a nearly identical mean age—while the remaining participants were assigned to 'Healthy Group 1'. Figure 4 presents the differences in demographic information and HAMD scores between the participants retained after data quality control and the original sample.

**Table2.**

Details for participants involved with data analysis

| Groups | Gender | Age (mean±SD) | HAMD (mean±SD) | HAMA (mean±SD) | BDI (mean±SD) | SAS (mean±SD) | Medication |
|---|---|---|---|---|---|---|---|
| MDD_1 (HAMD>16) | 7M, 10F | 28.18±11.72 | 22.94±5.74 | 25.41±7.27 | 23.65±10.62 | 56.24±11.63 | SSRI:8, SNRI:6 Others:3 |
| MDD_2 | 10M, 8F | 36.67±10.81 | 9.44+3.79 | 11.44±6.10 | 7.39±5.00 | 40.50±7.91 | SSRI:8 |
| (HAMD⩽16) | | | | | | | SNRI:2<br>Others:8 |
| Healthy Group 1(HC_1) | 7M,15F | 22.55±2.26 | / | / | / | / | / |
| Healthy Group 2(HC_2) | 6M, 11F | 35.18±7.13 | / | / | / | / | / |

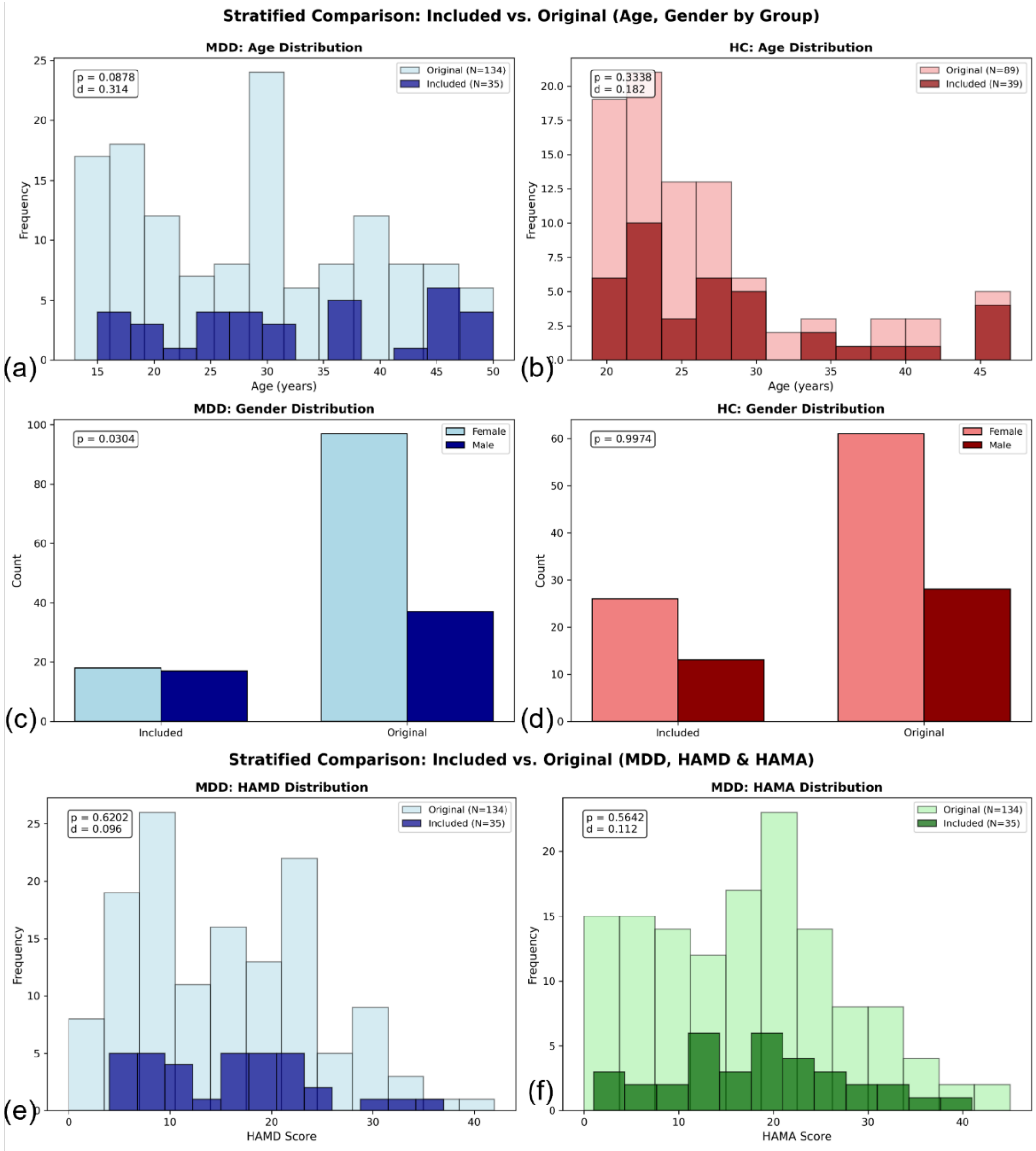

Fig 4. Demographic information, HAMD, and HAMA scores were compared between the original recruited participants and those retained after data preprocessing, separately for the MDD and HC groups. Among MDD patients, there was a significant gender difference between the original and retained samples (Figure c, $p < 0.05$). This observation is partly attributable to the poor near-infrared signal quality caused by long hair, as mentioned earlier. However, this gender difference was not observed in the healthy control group between recruited and retained participants. Therefore, in addition to long hair, other factors such as reduced vitality and poorer personal hygiene associated with MDD may also contribute to this discrepancy.

### 2.6 Resting-State EEG Local Maximum Response Calculation

Unlike previous studies that used deviant stimuli in the oddball paradigm to evoke activation—assuming individuals would show a P300 response at the Pz electrode and calculating the corresponding peak in hemodynamic coupling (Kennan et al., 2002; Phillips et al., 2016), which might bring difficulty with concentration in MDD patients, especially for those with greater severity. This study naturally focuses on the resting state. It calculated each individual's domain-specific maximum responses using global field power (GFP) as a substitute for task-induced activation.

Rather than calculating the GFP only once, GFP in this study was defined as the largest power moment within a sliding window compared to the 5 seconds before and after the given point. For each epoch, the top 5 peak values were extracted for consistency analysis.

This approach avoids the issue of inattention to externally induced stimuli (as in the oddball paradigm), which is particularly common in special populations such as children (Peisch et al., 2021) or patients with major depressive disorder or cognitive impairment (Tripathi et al., 2015)—especially those with greater severity—who often fail to attend to deviant stimuli due to habituation or cognitive overload (Pan et al., 2019). Furthermore, using the resting state helps prevent localized hemodynamic responses from reaching saturation due to sustained stimulation, which can lead to non-linear increases and interfere with experimental observation(Kennan et al., 2002; Phillips et al., 2016). In addition, it also avoids the potential issue that participant heterogeneity may render some event-related activities difficult to detect in certain individuals.

At the conclusion level, the calculation of local peak responses also improves measurement reliability by increasing the number of individual peaks considered, which helps balance error.

To focus the analysis on NVC, GFP of the EEG was only calculated over the regions covered by NIRS channels. This means the analysis only reflected activation differences from the prefrontal cortex, the frontoparietal junction, and the temporal lobes. The following EEG channels were excluded from the GFP calculation: 'Pz', 'P3', 'P4', 'P7', 'P8', 'PO3', 'PO4', 'Oz', 'O1', 'O2' when calculating the GFP.

GFP was calculated using the following formula:

$$GFP(t) = \sqrt{\sum_{i=1}^{N} (V_i(t) - \bar{V}(t))^2}$$

where N means the number of channels. $V_i(t)$ means the amplititude of the i-nd channels at the t moment. $\bar{V}(t)$ means the average amplititude of all channels at the t moment.

Each subject's data was segmented into 4 epochs (two eyes-open resting-state segments and two eyes-closed segments) for the calculation of local maximum GFP shown in *Fig.5*.

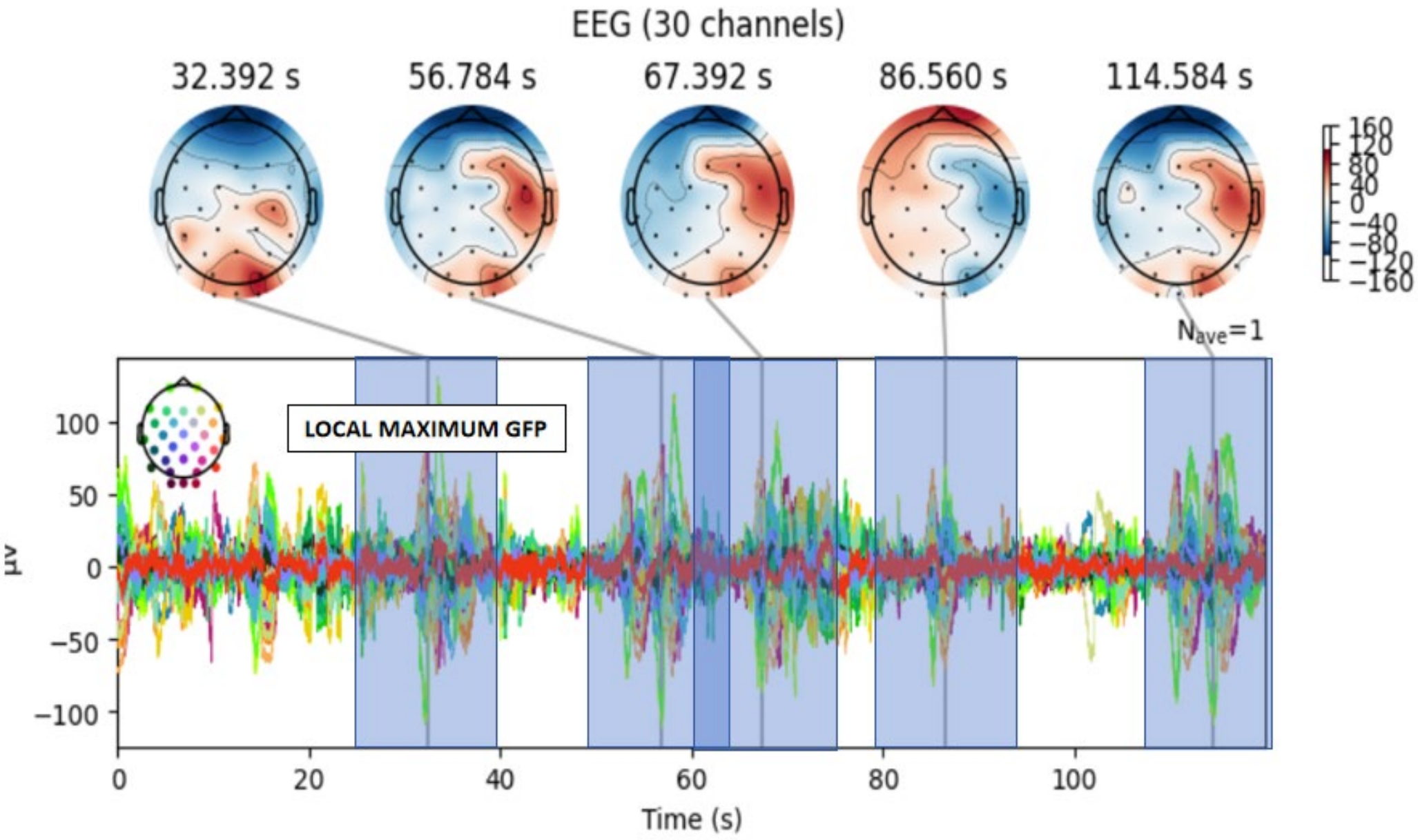

Fig.5.1. shows an example of the windows where the maximum GFP values occur within a single epoch. Each selected GFP peak is ensured to be greater than any GFP value in the preceding and following 5 seconds. Finally, the top 5 GFP peaks (in terms of amplitude) within each epoch were selected for further analysis.

**2.7 Neurovascular Coupling Calculation**

This study investigated changes in total hemoglobin (HbT) signals within an 10-second window following each EEG GFP peak. Specifically, for each peak detected in the GFP signal across different epochs, the spearman correlation coefficient was calculated between the EEG activation values and the HbT signal over the subsequent 10 seconds.

In this study, this correlation coefficient was defined as the Neurovascular Coupling Consistency Response Coefficient ($\boldsymbol{NVC_R}$). The maximum value of this coefficient within the observation window was regarded as the $mNVC_R$, and the time point at which this maximum occurred was defined as the $mNVC_RT$. Moreover, this value must be a peak, which was identified as the local maximum at time t, where the correlation coefficient $NVC_R(t)$ was strictly greater than both $NVC_R(t-1)$ and $NVC_R(t+1)$ .

Because the physical locations of fNIRS channels cannot exactly overlap with EEG electrode positions, it was necessary to perform interpolation on the full-head EEG topography before calculating the consistency response coefficient. This interpolation allowed the estimation of EEG activation values at locations equivalent to the fNIRS observation positions. EEG data was interpolated instead of fNIRS data, as the spatial resolution of fNIRS is higher. Preserving the spatial details of the fNIRS signal thus contributes to a more accurate assessment of the relationship between neural and hemodynamic activities.

Besides the NVC, a pronounced negative correlation between neural activity and hemodynamic signals emerged at the first few second at the beginning of the maximum response. This phenomenon is considered the "initial dip" of hemodynamic activity (Hu & Yacoub, 2012), and was defined in this study as: for individuals with a reasonable neurovascular coupling effect $(\mathrm{mNVC_R} > 0)$, the initial dip corresponds to the moment before the NVC peak where the correlation coefficient $\mathrm{r} < 0$ and reaches its most negative value (largest absolute magnitude).

And the minimum consistency was abbreviated as minimum initial dip value ($mID_R$), where the moment was called miniumun initial dip value time point ($mID_RT$).

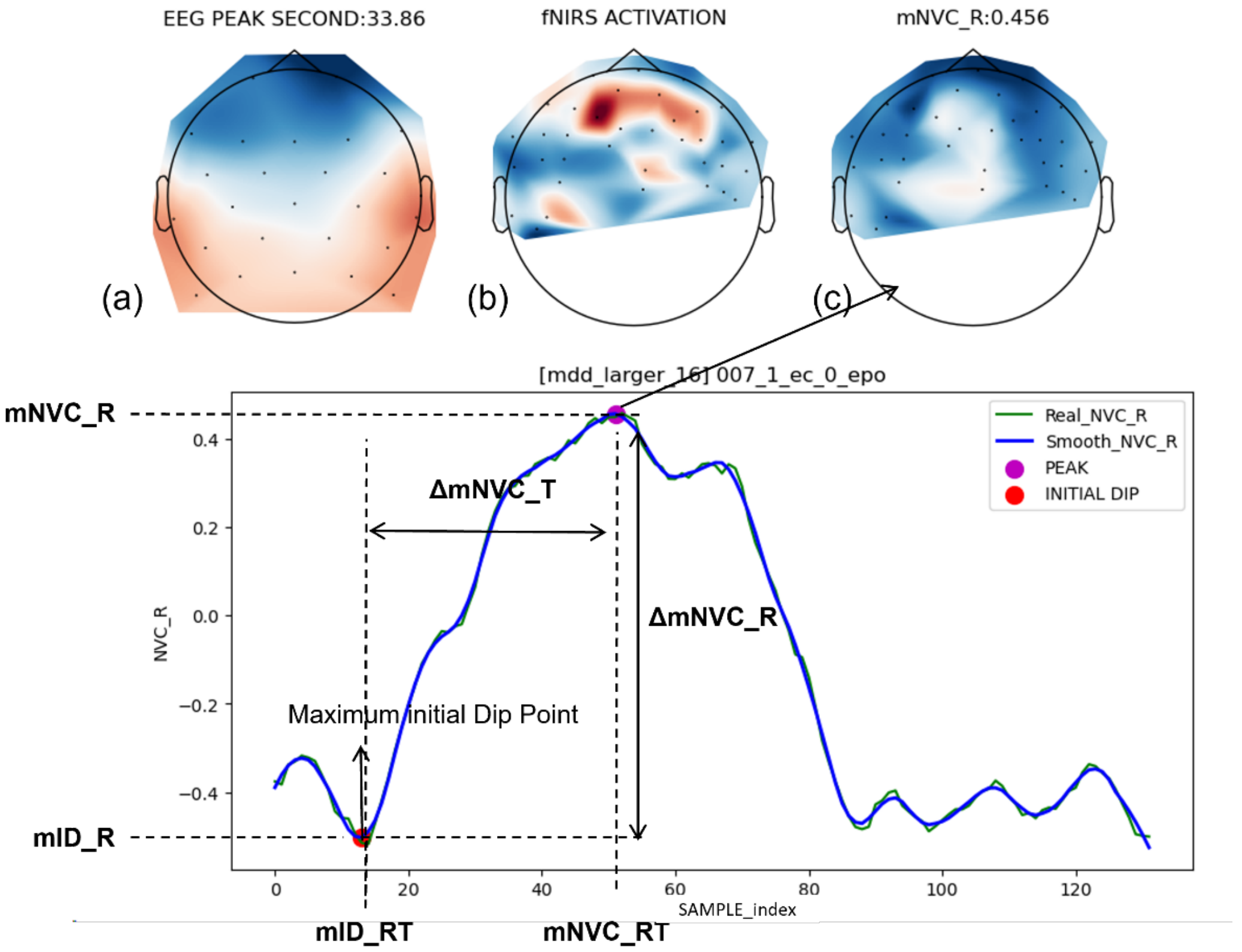

Fig.5.2. The example of the maximum neurovascular coupling consistency response coefficient (mNVC_R) and 10s (110 samples depends on the fNIRS frequency) of NVC_R after the peak GFP smoothed by Savitzky-Golay filter. As shown above, in this window, this GFP peak appears at the second 33.86 of the entire 120 second resting. Then after about 50 sample points, the NVC value reach its peak. We marked the corresponding Y value to mNVC_R, and the X value to mNVC_RT. And the minimum initial dip point appears about 17 sample point after the peak. (a) The corresponding activation of EEG at the GFP peak moment. (b) The corresponding activation of fNIRS at the GFP peak moment. (c) The activation mode of mNVC moment.

**2.8 Statics Analysis**

All statistical analyses were performed using Python. For each participant, repeated measurements (e.g., multiple epochs or recordings) were first averaged to obtain a single subject-level value before group-level statistical comparisons. Descriptive statistics were calculated as mean ± standard deviation (SD) for each group.

Differences among multiple groups were assessed using one-way analysis of covariance (ANCOVA) with Age and Gender as covariates to control for potential confounding effects of demographic variables. The F statistic and corresponding p value were calculated to evaluate the overall group effect after adjusting for these covariates. Effect sizes were quantified using partial eta squared (partial $\eta^2$), calculated from the ANCOVA results. When a significant or trend-level group effect($p<0.1$) was observed, post hoc pairwise comparisons were performed using estimated marginal means (controlling for Age and Gender at their mean values) with independent two-sample t-tests. To control for multiple comparisons, p values from pairwise tests were corrected using the Benjamini–Hochberg false discovery rate (FDR) procedure.

Adjusted p values < 0.05 were considered statistically significant.

For correlation analyses within the MDD patient group, partial correlation coefficients were calculated between NVC metrics and clinical symptom severity scales (HAMD, HAMA) while controlling for Age, Gender, and Medication status, allowing assessment of symptom-NVC relationships independent of these potential confounders.

Post-hoc statistical power was calculated for significant ANCOVA findings using the observed effect sizes (partial $\eta^2$) and sample sizes, with alpha set at 0.05. Power calculations were performed by converting observed F-statistics to Cohen's f effect sizes using the formula $f = \sqrt{(\eta^2/(1-\eta^2))}$. This approach allows readers to evaluate whether marginally significant or non-significant results reflect insufficient statistical power versus the absence of a true effect.

## 3. Results

After preprocessing the EEG and fNIRS data, a total of 17 MDD patients with HAMD larger than 16, 18 MDD patients with HAMD not larger than 16, and 22 healthy controls below 28 years old and 17 above 28 were included in the final analysis. The overall data rejection rate approached 65%, primarily because the MDD participants generally had long and poorer hair cleanliness, which might be related to their low vitality.

### 3.1 Age is a key factor influencing mNVC_R, yet MDD attenuates this relationship.

After controlling for Age and Gender using ANCOVA, a statistically significant group difference in mNVC_R was observed in the Eyes-Open condition *($F(3,64)=2.78$, $p=0.048$, partial $\eta^2=0.115$)*. Post-hoc comparisons using estimated marginal means revealed that mNVC_R was significantly higher in HC_2 *(0.35±0.07)* than in HC_1 *(0.27±0.10, $p_{FDR}=0.013$, Cohen's $d=0.90$)*. Additionally, a marginal trend was observed between MDD_1 *(0.30±0.10)* and HC_2 *($p_{uncorrected}=0.083$, Cohen's $d=0.57$)*. This age-related enhancement in NVC consistency was condition-specific: no significant group differences emerged in the Eyes-Closed condition *($F(3,64)=0.62$, $p=0.604$)*.

Age-related correlation analysis revealed that in healthy controls, mNVC_R showed a positive but slight trend with age in the Eyes-Open condition ($r=0.253$, $p=0.130$), suggesting age-related maturation of neurovascular coupling consistency as shown in *Fig.6*. However, this relationship was absent in MDD patients ($r=-0.063$, $p=0.725$), indicating that depression disrupts the typical age-related enhancement of NVC.

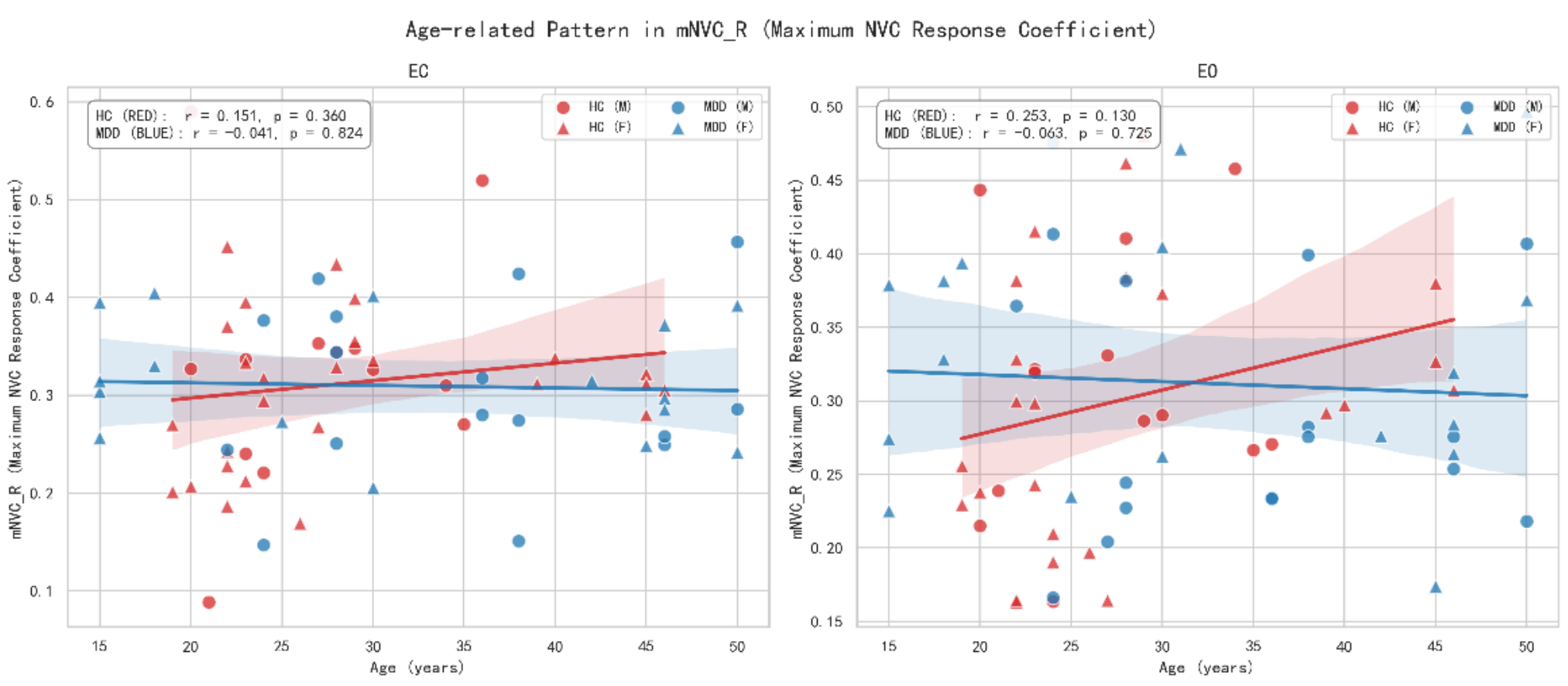

Fig 6: Association between mNVC_R and age in MDD patients and HC (without controlling for other factors). The left panel shows the eyes-closed condition, and the right panel shows the eyes-open condition. Circles represent male participants, and triangles represent female participants.

Notably, although patients in the MDD_1 group were generally younger (mean age 28.18 years), the difference in mNVC_R between MDD_1 and HC_2 or even MDD_2 did not reach statistical significance after FDR correction *(with HC_2 p_FDR=0.167, with MDD_2 p_FDR=0.665)*, suggesting that depression may attenuate the typical age-related increase in neurovascular coupling consistency.

No significant group differences were found in mNVC response latency (mNVC_RT) in either condition *(EC: F(3,64)=0.10, p=0.96; EO: F(3,64)=0.10, p=0.96),* likely due to high individual variability.

### 3.2 No significant group differences in the Maximum Initial Dip response coefficient (mID_R)

The initial dip, characterized by a transient negative correlation between neural activity and hemodynamic response during the oxygen consumption phase, showed no significant group differences after controlling for Age and Gender. In the Eyes-Closed condition, *F(3,60)=0.10, p=0.96*; in the Eyes-Open condition, *F(3,64)=1.44, p=0.24*. No significant pairwise differences were observed in mID_R or mID_RT after FDR correction.

Age-related correlation analysis showed that in MDD patients, mID_R exhibited a marginal positive trend with age in the Eyes-Closed condition (r=0.335, p=0.081), suggesting that older MDD patients may show deeper initial oxygen consumption. This relationship was not observed in healthy controls (r=-0.116, p=0.483).

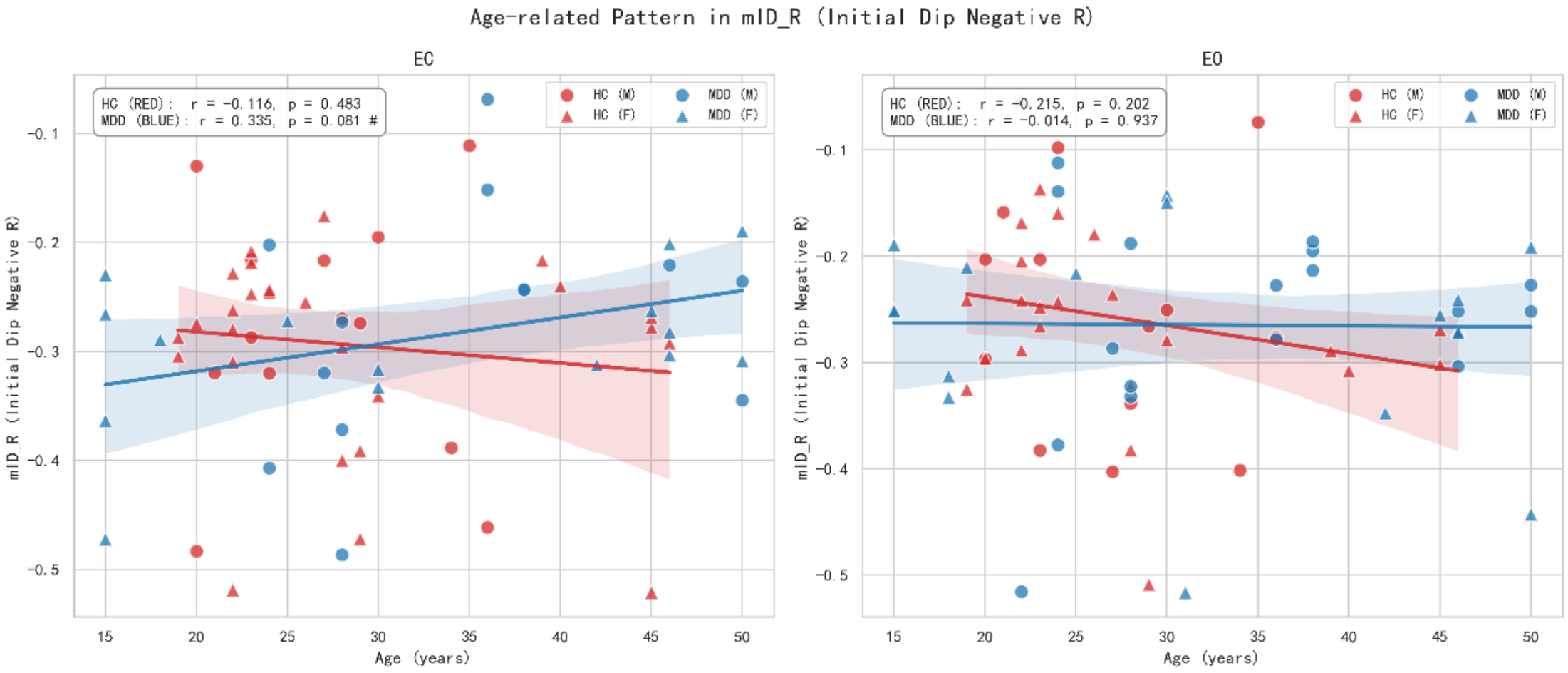

Fig 7: Association between mID_R and age in MDD patients and HC (without controlling for other factors). The left panel shows the eyes-closed condition, and the right panel shows the eyes-open condition. Circles represent male participants, and triangles represent female participants.

### 3.3 Depression is associated with an abnormally shortened duration of oxyhemoglobin replenishment

The time interval from the moment of maximum referred to as the mID_RT to the mNVC_RT defined as the **△mNVC-T** (delta maximum neurovascular coupling time interval), representing the duration of the hemodynamic replenishment phase. And the delta correlation value from the

mID_R to the mNVC_R was defined as **△mNVC-R** (delta maximum neurovascular coupling difference in correlation coefficient).

ANCOVA controlling for Age and Gender revealed a statistically significant group effect for delta-mNVC_R (the correlation change during replenishment) in the Eyes-Open condition *(F(3,64)=3.40, p=0.023, partial* $\eta^2$*=0.138).* Post-hoc analysis showed that HC_2 *(0.65±0.16)* exhibited significantly higher values compared to HC_1 *(0.47±0.18, p_FDR=0.004, Cohen's d=1.07)*. Marginal trends were observed between MDD_1 (*0.54±0.17*) and HC_2 *(p_uncorrected=0.049, Cohen's d=0.64)*, and between MDD_2 *(0.56±0.11)* and HC_2 *(p_uncorrected=0.052, Cohen's d=0.63).*

Age-related correlation analysis revealed a marginal positive trend between △mNVC_R and age in healthy controls during the Eyes-Open condition *(r=0.314, p=0.059)* as shown in Fig.9, indicating that older healthy individuals show greater correlation enhancement during hemodynamic replenishment. This age-related pattern was disrupted in MDD patients *(r=0.040, p=0.823)*.

For delta mNVC_T (time interval), a marginal positive correlation with age was observed in MDD patients during the Eyes-Closed condition *(r=0.345, p=0.072)* as shown in Fig.10, suggesting that older MDD patients may require longer replenishment time. In contrast, healthy controls showed no such relationship *(r=0.067, p=0.685)*.

This contribution was primarily driven by the Eyes-Open condition; no significant group differences emerged in the Eyes-Closed condition *(F(3,60)=0.44, p=0.72)*. No significant group differences were observed in delta-mNVC_R in either condition *(EC: F(3,60)=1.67, p=0.18; EO: F(3,64)=0.26, p=0.86)*.

These results indicate that age enhances the correlation consistency during the hemodynamic replenishment phase in healthy individuals. However, MDD patients showed attenuated values compared to age-matched healthy controls, suggesting that depression disrupts this age-related maturation of neurovascular coupling dynamics.

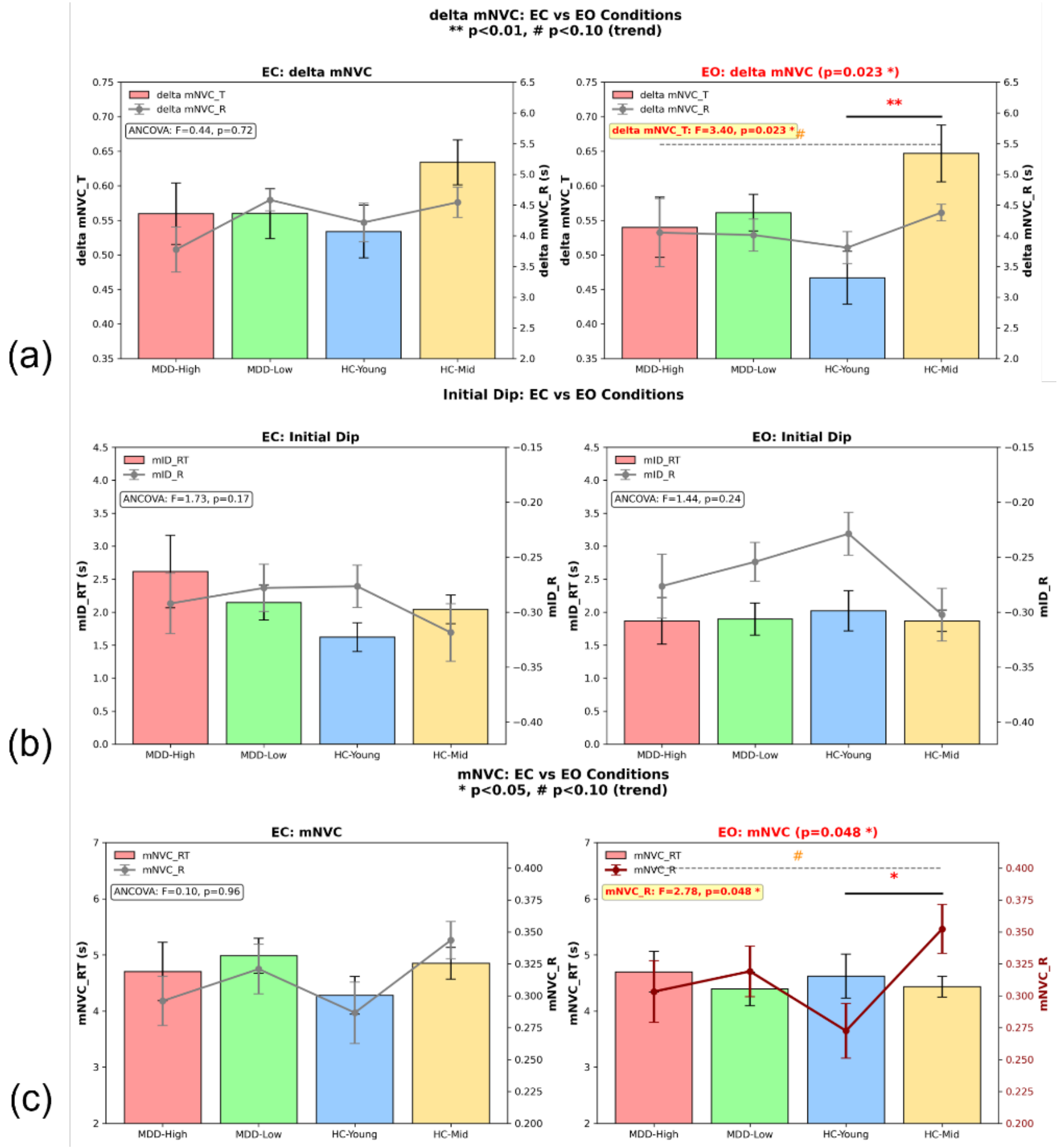

Fig.8. Bar and line charts were used to respectively present the time-related values and correlation coefficients under different definitions in different conditions. In (a), the bar chart shows the mNVC_T values across different groups, while the line chart illustrates the corresponding mNVC_R values. In Figure (b), the bar chart displays the mID_RT values for each group, and the line chart shows the corresponding mID_R values. In Figure (c), the bar chart represents the delta mNVC-T values across groups, while the line chart shows the delta mNVC-R values. It can be observed that group differences in time-related measures are minimal, whereas group differences in the consistency of neurovascular responses are more pronounced.

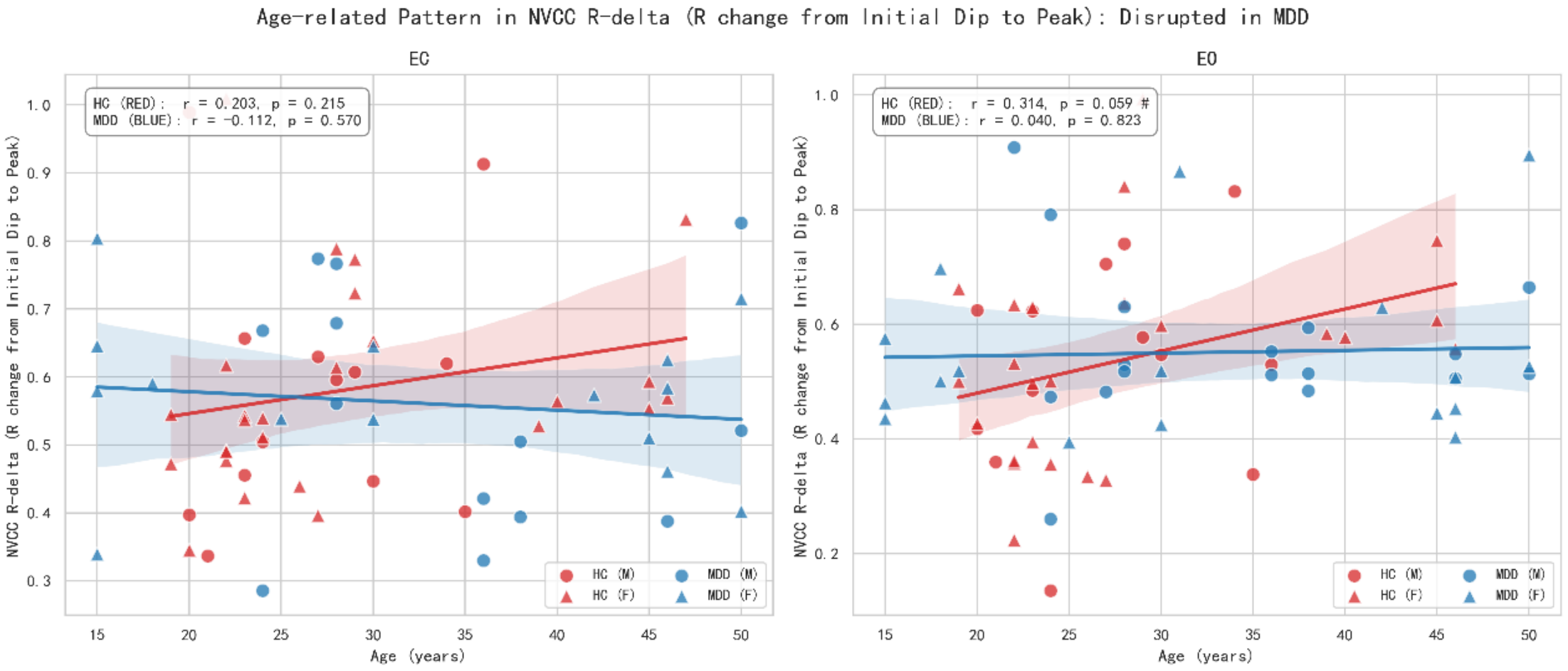

Fig 9. Association between ΔmNVC_R and age in MDD patients and HC (without controlling for other factors).

The left panel shows the eyes-closed condition, and the right panel shows the eyes-open condition. Circles represent male participants, and triangles represent female participants.

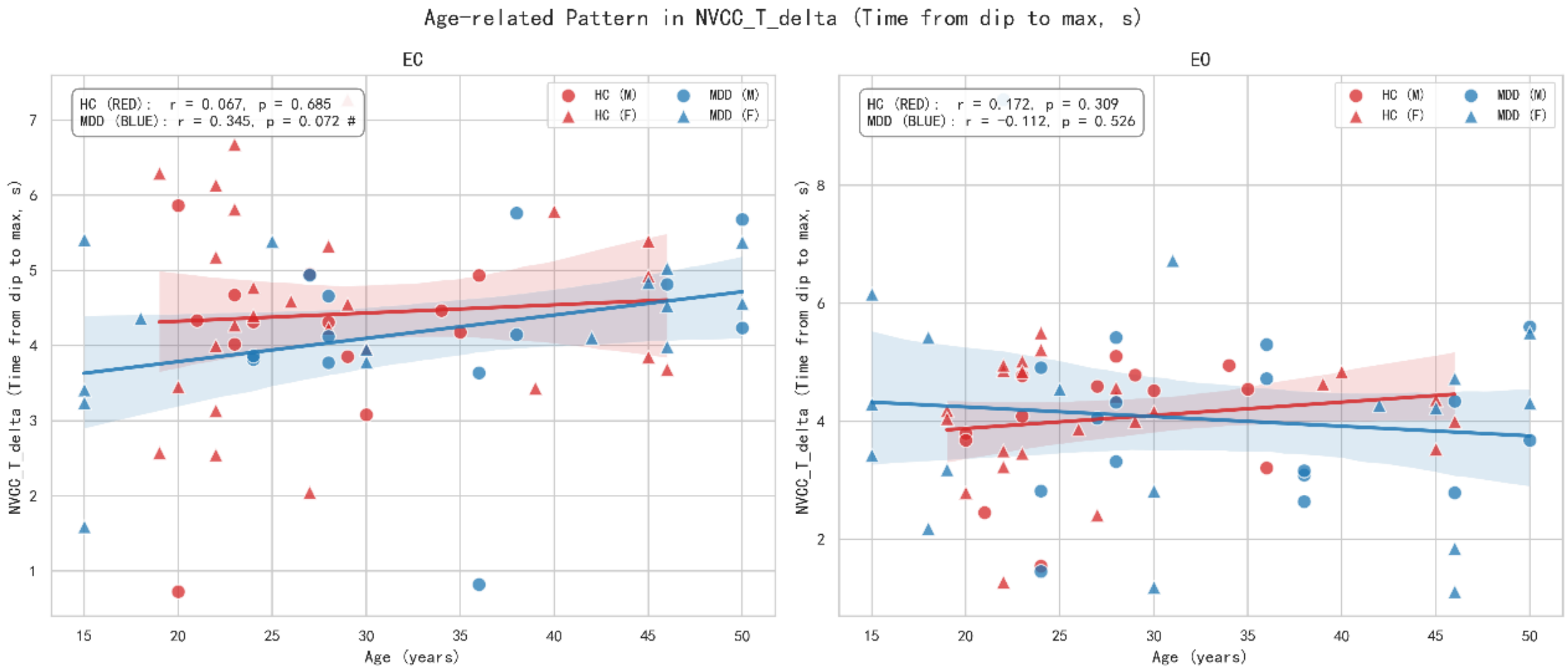

Fig 10: Association between Δ mNVC_RT and age in MDD patients and HC (without controlling for other factors). The left panel shows the eyes-closed condition, and the right panel shows the eyes-open condition. Circles represent male participants, and triangles represent female participants.

### 3.4 Partial correlation between NVC metrics and clinical symptoms in MDD

Within the MDD patient group, we examined the relationship between NVC metrics and clinical symptom severity while controlling for Age, Gender, and Medication status. In the Eyes-Closed condition, mNVC_R showed a trend-level negative partial correlation with HAMD scores *(partial r = -0.34, p = 0.060)*, suggesting that more severe depressive symptoms were associated with lower neurovascular coupling consistency, independent of demographic factors and medication effects. A similar trend was observed in the Eyes-Open condition *(partial r = -0.29, p = 0.095)* as shown in Fig.11.

For the initial dip metrics, mID_R showed a marginal positive partial correlation with HAMA anxiety scores in the Eyes-Closed condition *(partial r = 0.318, p = 0.100) as shown in Fig.12*, suggesting that higher anxiety may be associated with deeper oxygen consumption during neural activation.

For delta mNVC metrics, delta mNVC_R showed marginal negative partial correlations with *HAMA (partial r = -0.296, p = 0.090)* in the Eyes-Open condition. Similarly, the delta mNVC_T showed marginal negative partial correlations as shown in Fig 13, both indicating that greater symptom severity may be associated with reduced correlation enhancement during hemodynamic replenishment. These relationships remained after controlling for age, gender, and medication effects.

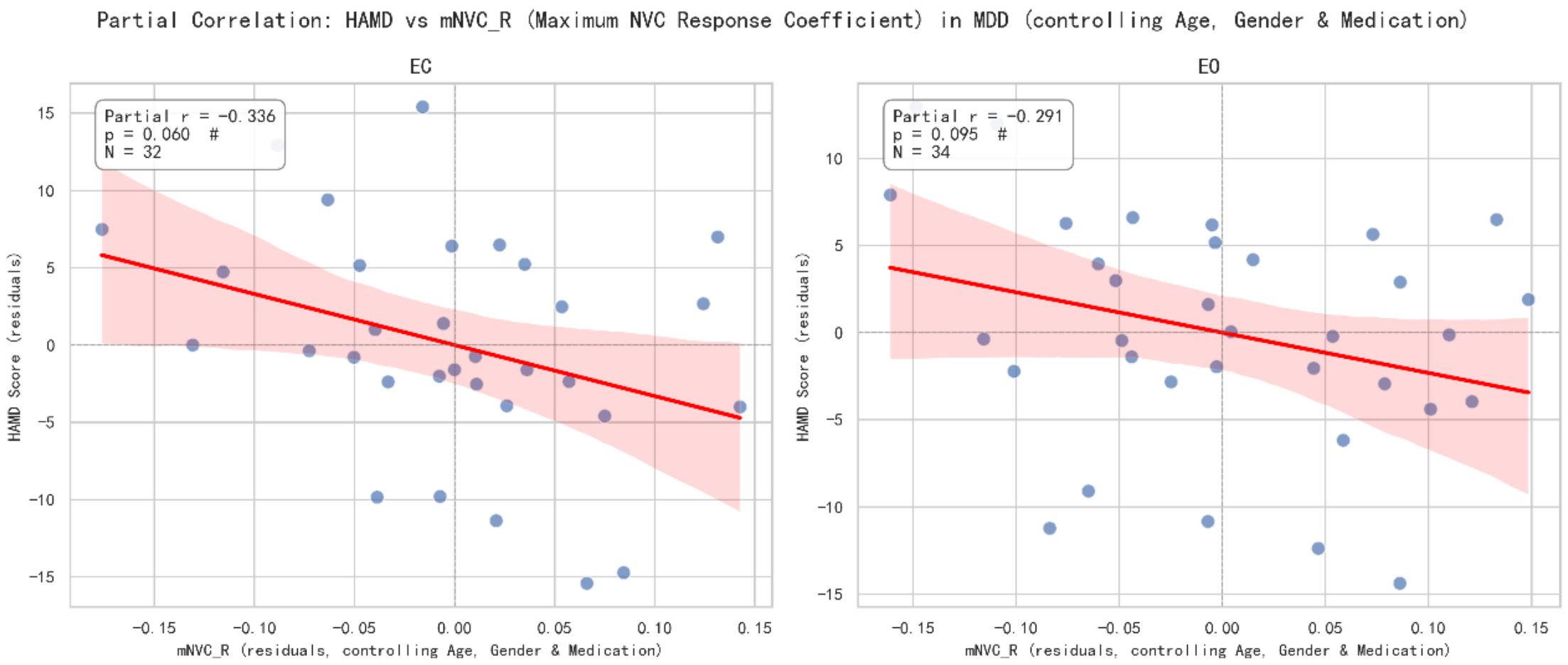

Fig 11: The association between mNVC_R and HAMD scores in MDD patients after controlling for sex, age, and medication (SSRI, SNRI, Others).The left panel shows the eyes-closed condition, and the right panel shows the eyes-open condition.

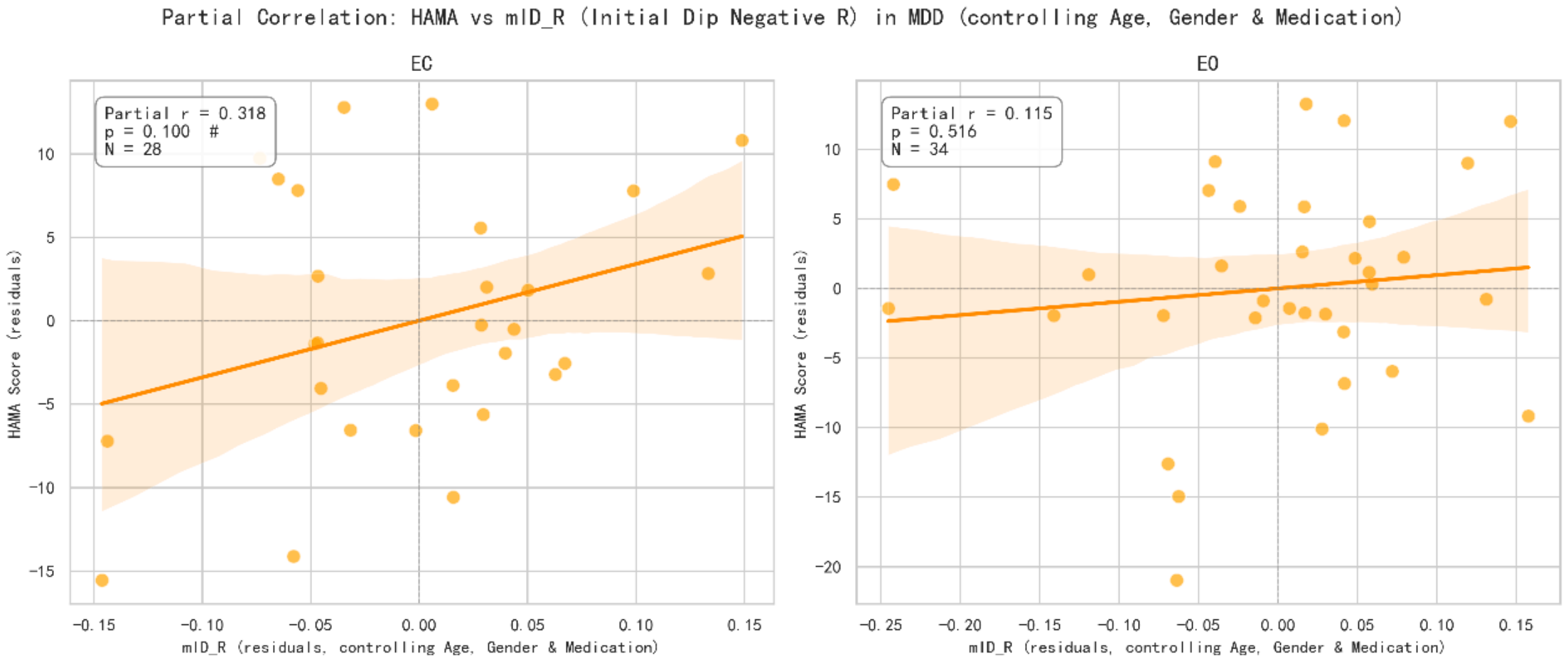

Fig 12: The association between mID_R and HAMA scores in MDD patients after controlling for sex, age, and medication (SSRI, SNRI, Others). The left panel shows the eyes-closed condition, and the right panel shows the eyes-open-condition.

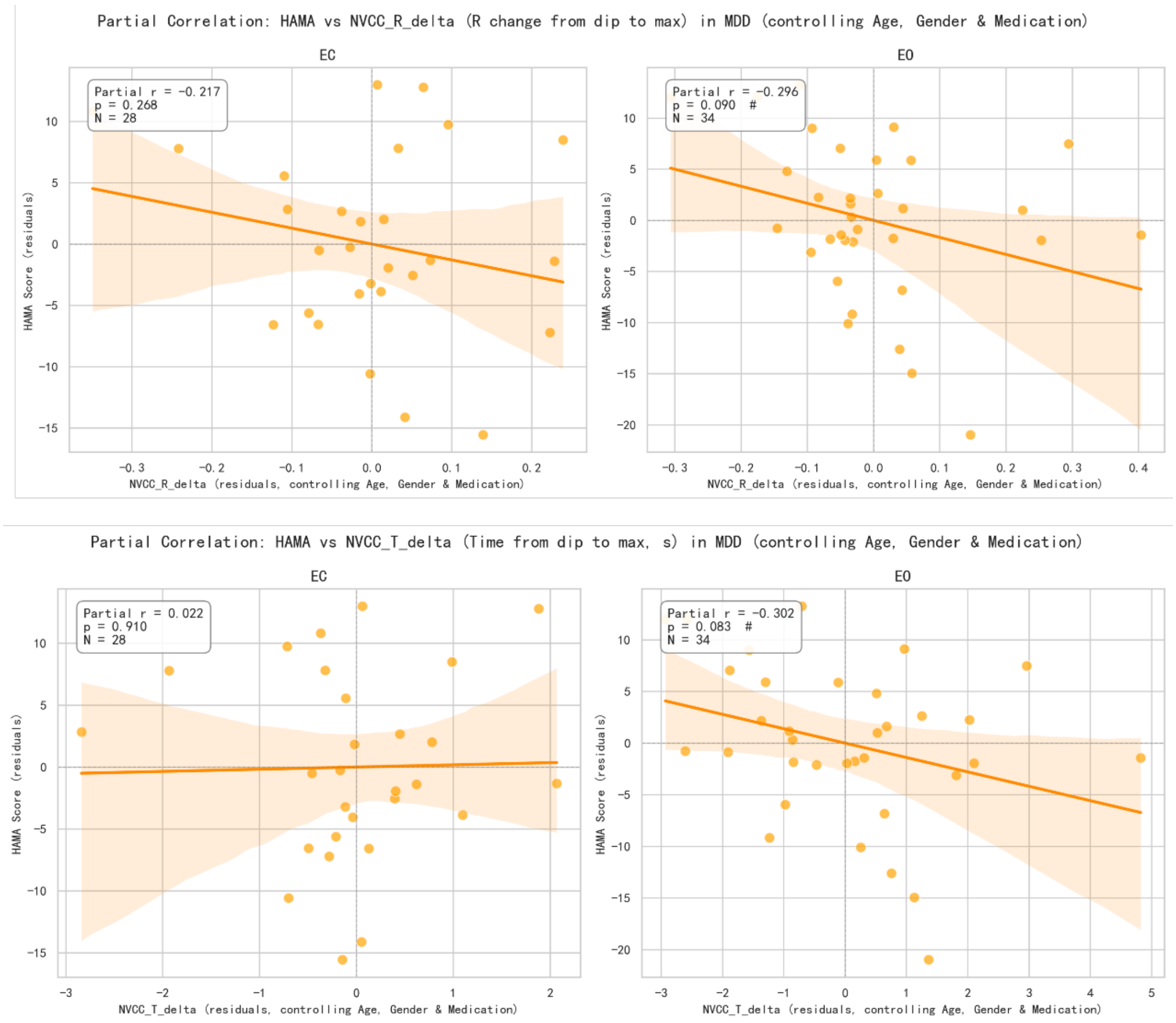

Fig 13: (a) The association between delta-mNVC_R and HAMA scores in MDD patients after controlling for sex, age, and medication (SSRI, SNRI, Others). (b) The association between delta-mNVC_T and HAMD scores in MDD patients after controlling for sex, age, and medication (SSRI, SNRI, Others). The left panel shows the eyes-closed condition, and the right panel shows the eyes-open condition.

## 4. Discussion

In the present study, we utilized simultaneous EEG-fNIRS recording to investigate neurovascular coupling (NVC) in the prefrontal cortex among healthy controls and patients with major depressive disorder (MDD) with different severity. To our knowledge, this is the first study to examine severity-related differences in resting-state NVC using this multimodal approach while controlling for age and gender effects through ANCOVA.

We recruited 206 participants and recorded two-minute resting-state sessions under both eyes-open and eyes-closed conditions using simultaneous EEG-fNIRS. After rigorous quality control, 74 participants (17 MDD subjects with HAMD>16, 18 MDD subjects with HAMD ⩽16, 22 younger healthy controls <28 years, and 17 older healthy controls ≥28 years) were retained for analysis. The 65% exclusion rate, primarily driven by poor fNIRS signal quality due to hair coverage, represents a significant technical limitation that we address below.

Our ANCOVA results, controlling for age and gender, revealed that older healthy individuals exhibited slightly significantly enhanced neurovascular coupling consistency (mNVC_R) compared to younger healthy controls (p=0.013, Cohen's d=0.90), indicating age-related maturation of NVC. However, this age-related enhancement was attenuated in MDD patients showing intermediate values that did not differ significantly from either healthy control subgroup. Additionally, we observed that the correlation change during hemodynamic replenishment (ΔmNVC_T) was significantly greater in older healthy controls compared to younger controls (p=0.004, Cohen's d=1.07), with MDD patients again showing attenuated responses. These effects were primarily observed during the eyes-open condition, suggesting that arousal state modulates the expression of NVC alterations in depression..

**4.1 Interpretation of neurovascular coupling alterations in MDD**

The observed attenuation of age-related NVC enhancement in MDD patients may reflect disrupted neurovascular plasticity. Neurovascular coupling consistency (mNVC_R) likely reflects the precision and reliability of activity-dependent vascular responses, which mature with age in healthy individuals but appear compromised in depression. The trend-level negative correlation between mNVC_R and HAMD scores (partial r=-0.34, p=0.060, controlling for age, gender, and medication) supports the interpretation that more severe depressive symptoms are associated with poorer neurovascular coupling, independent of demographic and treatment factors.

The finding that ΔmNVC_T (correlation change during replenishment) follows a similar pattern—enhanced in older healthy individuals but attenuated in MDD—suggests that depression may impair the hemodynamic recovery phase following neural activation. This could reflect either primary vascular dysfunction or secondary changes due to chronic illness. Previous studies have demonstrated that depression is associated with endothelial dysfunction (Feng et al., 2024) and reduced vascular elasticity, which could mechanistically explain the observed NVC alterations.

Our EEG topographic analysis (Fig 14) revealed distinct prefrontal activation patterns in the MDD_2 group; however, these differences in neural activity alone did not account for the observed NVC alterations, suggesting that the primary deficit lies in the neurovascular interface rather than in neural activity per se. This interpretation is consistent with emerging evidence that psychiatric disorders involve not only neural circuit dysfunction but also supporting vascular and metabolic systems (Banasr et al., 2021; Zhang et al., 2018). Recent microstate research in drug-naïve MDD has revealed severity-dependent temporal dynamics, anxiety-related microstate alterations, and sex-dependent heterogeneity in microstate patterns(He et al., 2026; Zhang et al., 2026; Zhang et al., 2025). These complementary findings suggest that transient neural states captured at GFP peaks may represent functionally relevant moments when neural-hemodynamic coupling is maximally engaged. The convergence between microstate abnormalities and our observed NVC disruptions in MDD reinforces the notion that depression involves alterations in both the temporal dynamics of neural states and their associated metabolic/vascular responses. Future studies combining microstate analysis with simultaneous hemodynamic recording could help clarify whether specific microstate configurations are associated with distinct NVC patterns, potentially revealing state-dependent neurovascular dysfunction in depression.

**4.2 The confounding role of medication.**

A critical limitation of this study is the inability to fully disentangle medication effects from disease-phase effects. As shown in Table 2, the majority of MDD patients were receiving antidepressant treatment (MDD_1: 94% medicated; maintenance-phase: 83% medicated), with SSRIs and SNRIs as the predominant classes. Previous research suggests that SSRIs may enhance endothelial function and vasodilation (Delialis et al., 2022), which could theoretically influence NVC metrics.

However, several observations argue against medication being the primary driver of our findings. First, our partial correlation analyses controlling for medication status still revealed trend-level associations between symptom severity and NVC metrics. Second, if SSRIs were improving neurovascular coupling, we would expect the MDD_2 group (with longer treatment duration) to show values closer to healthy controls, yet we observed continued attenuation of the age-related NVC pattern.

Nevertheless, we cannot definitively rule out medication effects with the current cross-sectional design. The medication distribution was not balanced between groups (by necessity, as treatment status defines clinical management), making statistical "control" of medication incomplete. Future longitudinal studies tracking patients from drug-naïve acute phase through treatment and recovery would be necessary to fully parse medication effects from disease trajectory effects..

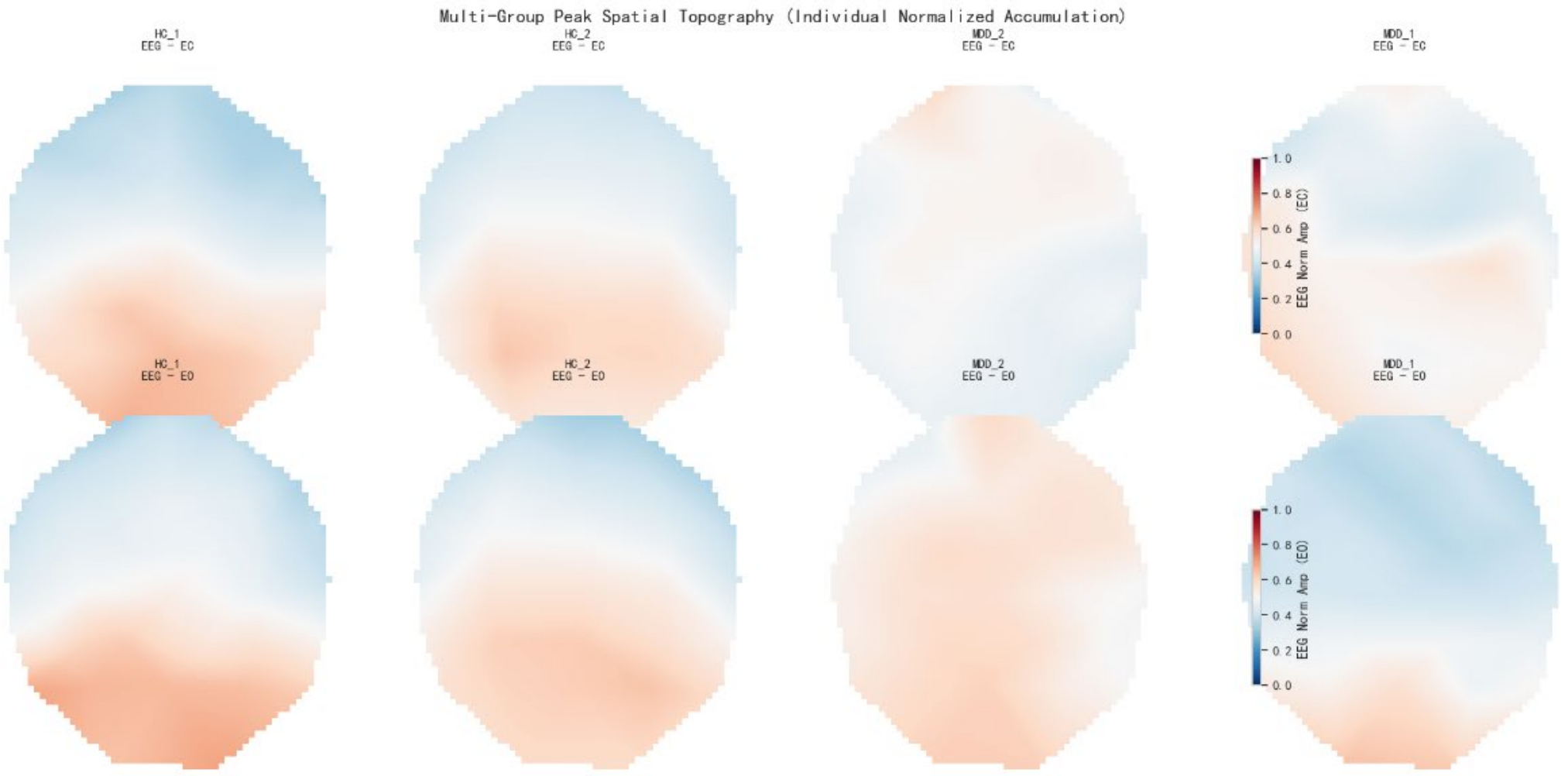

Fig.14. Average Topographical Maps of EEG Peaks in different groups across different conditions

**4.3 Condition-specific effects: Eyes-open versus eyes-closed**

Nearly all statistically significant group differences emerged specifically during the eyes-open condition, with minimal or absent effects during eyes-closed rest. This pattern likely reflects the role of arousal and attentional state in modulating neurovascular coupling. Previous EEG studies have shown that depression patients exhibit reduced alpha power modulation between eyes-open and eyes-closed states, suggesting difficulty transitioning between arousal levels. Our findings extend this observation to the neurovascular domain: the eyes-open state may provide a "stress test" that reveals latent NVC impairments not apparent during lower-arousal rest.

This condition-specificity has methodological implications for future NVC studies in psychiatric populations, suggesting that challenge paradigms or arousal manipulations may be more sensitive than passive rest for detecting neurovascular alterations.

**4.4 Statistical power and interpretation of marginal findings**

Post-hoc power analysis revealed that our significant finding for mNVC_R ($F=2.78$, $p=0.048$) achieved 64.5% observed power with a medium effect size (partial $\eta^2=0.115$, Cohen's $f=0.35$), falling short of the conventional 80% threshold. For delta mNVC_T ($F=3.40$, $p=0.023$), observed power was 74.2% (partial $\eta^2=0.138$, Cohen's $f=0.38$). These power estimates indicate that while we successfully detected statistically significant medium-sized effects, the study was underpowered to detect smaller effects and likely missed some true differences between MDD subgroups.

The small sample sizes (n=17-22 per group after 65% attrition) represent a major limitation. Several theoretically important comparisons—such as acute versus maintenance phase, or medication-stratified analyses—did not reach statistical significance, but this non-significance cannot be interpreted as evidence of no difference. The marginal findings ($0.05<p<0.10$) observed in several post-hoc comparisons likely reflect genuine medium effects that our sample size was insufficiently powered to detect reliably, rather than representing "trend-level" evidence requiring only slight replication to reach significance.

We therefore frame these findings as preliminary evidence that age-related neurovascular coupling maturation is disrupted in MDD, requiring replication in larger, adequately powered samples before firm clinical conclusions can be drawn.

**4.5 Selection bias and generalizability**

The 65% exclusion rate due to poor fNIRS signal quality raises concerns about selection bias. As documented in our analysis, the retained MDD_1 sample was significantly older (28.18 vs. 23.34 years, $p=0.088$) and had different gender distribution ($p=0.030$) compared to the originally recruited cohort. This age shift could theoretically bias our age-related interpretations; however, our stratified analysis showed that the magnitude of age-related selection bias did not differ substantially between MDD and healthy control groups (effect size difference <0.2), and all primary analyses statistically controlled for age as a covariate.

Nevertheless, the retained sample may not be fully representative of the broader MDD population. Participants with poorer self-care (often correlated with illness severity) were more likely to be excluded due to hair-related signal quality issues, potentially leading to underrepresentation of the most severe cases. This attrition pattern limits the generalizability of our findings, particularly to more severely ill or treatment-resistant populations.

**4.6 Cross-sectional design and causal interpretation**

Our study compared distinct groups (MDD_1,2 and HC_1,2) at a single time point. We cannot infer longitudinal recovery trajectories or establish causality from these cross-sectional comparisons. The observation that maintenance-phase patients show intermediate NVC values between acute patients and healthy controls is consistent with—but does not demonstrate—a recovery process. Alternative explanations include selection effects (patients with better baseline neurovascular function being more likely to achieve remission) or stable trait differences that pre-date illness onset.

Future longitudinal studies tracking individual patients from acute phase through recovery are needed to determine whether NVC normalization occurs with treatment response, and whether baseline NVC predicts treatment outcomes.

### 4.7 Limitations and future directions

Beyond the statistical power and selection bias issues discussed above, several methodological limitations warrant mention:

1. Multiple comparisons: We tested multiple NVC metrics (mNVC_R, mNVC_RT, mID_R, mID_RT, delta mNVC_R, delta mNVC_T) across two conditions (EC, EO) with FDR correction applied within each analysis family rather than across all tests. This approach balances Type I and Type II error risk but means our findings should be considered exploratory and hypothesis-generating pending replication.
2. Spatial specificity: Our fNIRS coverage was limited to prefrontal and temporal cortices. Whether NVC alterations extend to other brain regions relevant to depression (e.g., posterior cingulate, hippocampus) remains unknown.
3. Medication heterogeneity: Beyond the SSRI/SNRI distinction, patients varied in specific medications, dosages, and treatment duration. More fine-grained pharmacological data would strengthen future medication-effect analyses.
4. Clinical heterogeneity: The MDD groups likely included patients with different subtypes, comorbidities, and illness trajectories that could not be parsed with our sample size.

Future research should prioritize: (a) longitudinal designs tracking NVC changes with treatment; (b) drug-naïve patient samples or medication-washout protocols to isolate disease effects; (c) larger samples powered for medication-stratified and severity-stratified analyses; (d) improved fNIRS data quality protocols to reduce attrition; and (e) multisite replication to establish generalizability.

### 4.8 Clinical implications

Despite these limitations, our findings suggest that neurovascular coupling assessment may complement traditional neural-focused approaches to understanding depression pathophysiology. The observation that NVC alterations persist in maintenance-phase patients despite symptomatic improvement hints that vascular dysfunction may be a residual deficit requiring targeted intervention. Therapies enhancing vascular health—such as aerobic exercise, dietary interventions, or novel vasoactive compounds—merit investigation as adjunctive treatments for depression, though such recommendations remain speculative pending controlled trials.

Our resting-state multimodal EEG-fNIRS approach, once technical challenges are addressed, could potentially serve as a biomarker for monitoring treatment response or stratifying patients for vascular-targeted interventions, though substantial validation work would be required before clinical implementation.

# Conclusion

Using simultaneous EEG-fNIRS recording under resting conditions, we provide preliminary evidence that age-related neurovascular coupling maturation is disrupted in major depressive disorder, particularly during eyes-open arousal states. ANCOVA analyses controlling for age and gender revealed that older healthy individuals show enhanced NVC consistency and greater correlation change during hemodynamic replenishment compared to younger individuals, but this pattern is attenuated in MDD patients regardless of clinical severity. While medication effects cannot be definitively excluded, partial correlation analyses controlling for treatment status suggest symptom severity independently predicts NVC impairment. These findings, though limited by small sample size (64-74% statistical power) and high attrition (65% exclusion rate), position neurovascular dysfunction as a potential contributor to depression pathophysiology worthy of further investigation in larger, longitudinal samples.

**Data Availability Statement**

The preprocessed data and code supporting the conclusions of this article will be made available upon reasonable request. All the data will be shared after the whole collection project has been completed. We intend to release a portion of the data alongside another ongoing algorithmic project in the near future.

**Ethics Statement**

The studies involving human participants were approved by the Tsinghua University Ethics Committee (Approval No. THU01-20240104) and the Institutional Medicine Review Board of Beijing Huilongguan Hospital (approval number:2024-106-research, 2026-10-research).

All studies were conducted in accordance with local legislation and institutional requirements. Written informed consent was obtained from all participants.

**Code Availability**

All code used for data preprocessing and technical validation is available upon request, or when the project has been completed.

**Acknowledgements**

This work was supported by the Capital's Funds for Health Improvement and Research (2022-2-2133, 2024-1-2131) and Beijing Municipal High Rank Public Health Researcher Training Program (03-18).

**Author Contributions**

Conceptualization: XW, FY
Clinical Support: FY, ZW
Methodology: XW, FY
Investigation: ZW
Visualization: XW, ZY, SY
Supervision: ZW
Writing—Original Draft: XW, FY
Writing—Review & Editing: ZW, XW

**Competing Interests**

Authors declare that they have no competing interests.